\documentclass
[aps,prb,superscriptaddress,reprint,showpacs,amssymb,amsmath,floatfix,citeautoscript,longbibliography]{revtex4-1}

\usepackage{graphicx} 
\usepackage{dcolumn} 
\usepackage{bm} 

\usepackage{ulem}
\usepackage[]{color}
\newcommand{\red}[1]{\textcolor{red}{#1}}

\begin{document}

\title{Orientation of ground-state orbital in CeCoIn$_5$ and CeRhIn$_5$}

\author{M.~Sundermann}
 \affiliation{Institute of Physics II, University of Cologne, Z{\"u}lpicher Stra{\ss}e 77, 50937 Cologne, Germany}
  \affiliation{Max Planck Institute for Chemical Physics of Solids, N{\"o}thnitzer Stra{\ss}e 40, 01187 Dresden, Germany}
\author{A.~Amorese}
 \affiliation{Institute of Physics II, University of Cologne, Z{\"u}lpicher Stra{\ss}e 77, 50937 Cologne, Germany}
  \affiliation{Max Planck Institute for Chemical Physics of Solids, N{\"o}thnitzer Stra{\ss}e 40, 01187 Dresden, Germany}
\author{F.~Strigari}
 \altaffiliation{Present address: Bundesanstalt f{\"u}r Stra{\ss}enwesen, Cologne, Germany}
 \affiliation{Institute of Physics II, University of Cologne, Z{\"u}lpicher Stra{\ss}e 77, 50937 Cologne, Germany}
\author{B.~Leedahl}
	\affiliation{Max Planck Institute for Chemical Physics of Solids, N{\"o}thnitzer Stra{\ss}e 40, 01187 Dresden, Germany}
\author{L.~H.~Tjeng}
	\affiliation{Max Planck Institute for Chemical Physics of Solids, N{\"o}thnitzer Stra{\ss}e 40, 01187 Dresden, Germany}
\author{M.~W.~Haverkort}
  \affiliation{Institute for Theoretical Physics, Heidelberg University, Philosophenweg 19, 69120 Heidelberg, Germany}
\author{H.~Gretarsson}
  \affiliation{PETRA III, Deutsches Elektronen-Synchrotron (DESY), Notkestra{\ss}e 85, 22607 Hamburg, Germany}
	\affiliation{Max Planck Institute for Chemical Physics of Solids, N{\"o}thnitzer Stra{\ss}e 40, 01187 Dresden, Germany}
\author{H.~Yava\c{s}}
 \altaffiliation{Present address: SLAC National Accelerator Lab., 2575 Sand Hill Rd, Menlo Park, CA 94025, USA}
  \affiliation{PETRA III, Deutsches Elektronen-Synchrotron (DESY), Notkestra{\ss}e 85, 22607 Hamburg, Germany}
\author{M.~Moretti~Sala}
 \altaffiliation{Present address: Dipartimento di Fisica, Politecnico di Milano, Piazza Leonardo da Vinci 32, I-20133 Milano, Italy}
  \affiliation{European Synchrotron Radiation Facility, 71 Avenue des Martyrs, CS40220, F-38043 Grenoble Cedex 9, France}
\author{E.~D.~Bauer} 
  \affiliation{Los Alamos National Laboratory, New Mexico 87545, USA} 
\author{P.~F.~S.~Rosa} 
  \affiliation{Los Alamos National Laboratory, New Mexico 87545, USA} 
\author{J.~D.~Thompson} 
  \affiliation{Los Alamos National Laboratory, New Mexico 87545, USA} 
\author{A.~Severing}
  \affiliation{Institute of Physics II, University of Cologne, Z{\"u}lpicher Stra{\ss}e 77, 50937 Cologne, Germany}
  \affiliation{Max Planck Institute for Chemical Physics of Solids, N{\"o}thnitzer Stra{\ss}e 40, 01187 Dresden, Germany}	
\date{\today}

\begin{abstract}
We present core level non-resonant inelastic x-ray scattering (NIXS) data of the heavy fermion compounds CeCoIn$_5$ and CeRhIn$_5$ measured at the Ce $N_{4,5}$-edges. The higher than dipole transitions in NIXS allow determining the orientation of the $\Gamma_7$ crystal-field ground-state orbital within the unit cell. The crystal-field parameters of the Ce$M$In$_5$ compounds and related substitution phase diagrams have been investigated in great detail in the past; however, whether the ground-state wavefunction is the $\Gamma_7^+$ ($x^2\,-\,y^2$) or $\Gamma_7^-$ ($xy$ orientation) remained undetermined. We show that the $\Gamma_7^-$ doublet with lobes along the (110) direction forms the ground state in CeCoIn$_5$ and CeRhIn$_5$. A comparison is made to the results of existing DFT+DMFT calculations.  
\end{abstract}

\maketitle
\section{Introduction}
At high temperature, heavy-fermion materials are described by decoupled localized $f$ electrons and conduction electron bands. Upon cooling, the localized $f$ electrons start to interact with the conduction electrons ($cf$-hybridization) and become partially delocalized. The resulting entangled fluid consists of heavy quasiparticles with masses up to three orders of magnitude larger than the free electron mass. These quasiparticles may undergo magnetic or superconducting transitions. In the Doniach phase diagram temperature $T$ versus exchange interaction $\cal{J}$, magnetic order prevails for small $\cal{J}$ whereas a non-magnetic Kondo singlet state forms for large $\cal{J}$. Between these two regimes quantum critical behaviour occurs which is often accompanied by a superconducting dome that hides a quantum critical point (QCP).\cite{Hilbert2007,Wirth2016}  Understanding how these quasiparticles, that have atomic-like as well as itinerant character, give rise to these ground states is a challenging question in condensed-matter physics, and the answer to this question will provide predictive understanding of these quantum states of matter.\cite{Coleman2007} 

The tetragonal compounds Ce$M$In$_5$ ($M$\,=\,Co, Rh, Ir) are heavy fermion compounds that display different ground states for different transition metal ions; for $M$\,=\,Co and Ir the ground state is superconducting ($T_c$\,=\,2.3 and 0.4\,K) and for $M$\,=\,Rh it is antiferromagnetic ($T_N$\,=\,3.8\,K).\cite{Thompson2012} High-quality Ce$M$In$_5$ crystals can be grown, making this family suitable for determining the parameter that drives the different ground states. Within the above mentioned extended Doniach phase diagram, CeRhIn$_5$ is on the weak side of hybridization, CeCoIn$_5$ close to the QCP and CeIrIn$_5$ is on the side of stronger $cf$-hybridization, i.e.\,superconductivity goes along with stronger $cf$-hybridization. Although there are strong indications for localization (Rh) and delocalization (Co,Ir) in, e.g., the size of the Fermi surface,\cite{Haga2001,Fujimori2003,Harrison2004,Shishido2005,Settai2007,Shishido2007,Goh2008,Allen2013} it is not possible to detect the differences in $f$ occupations. They are so subtle that they are below the detection limit.\cite{Sundermann2016}

\begin{figure}[t]
    \includegraphics[width=0.8\columnwidth]{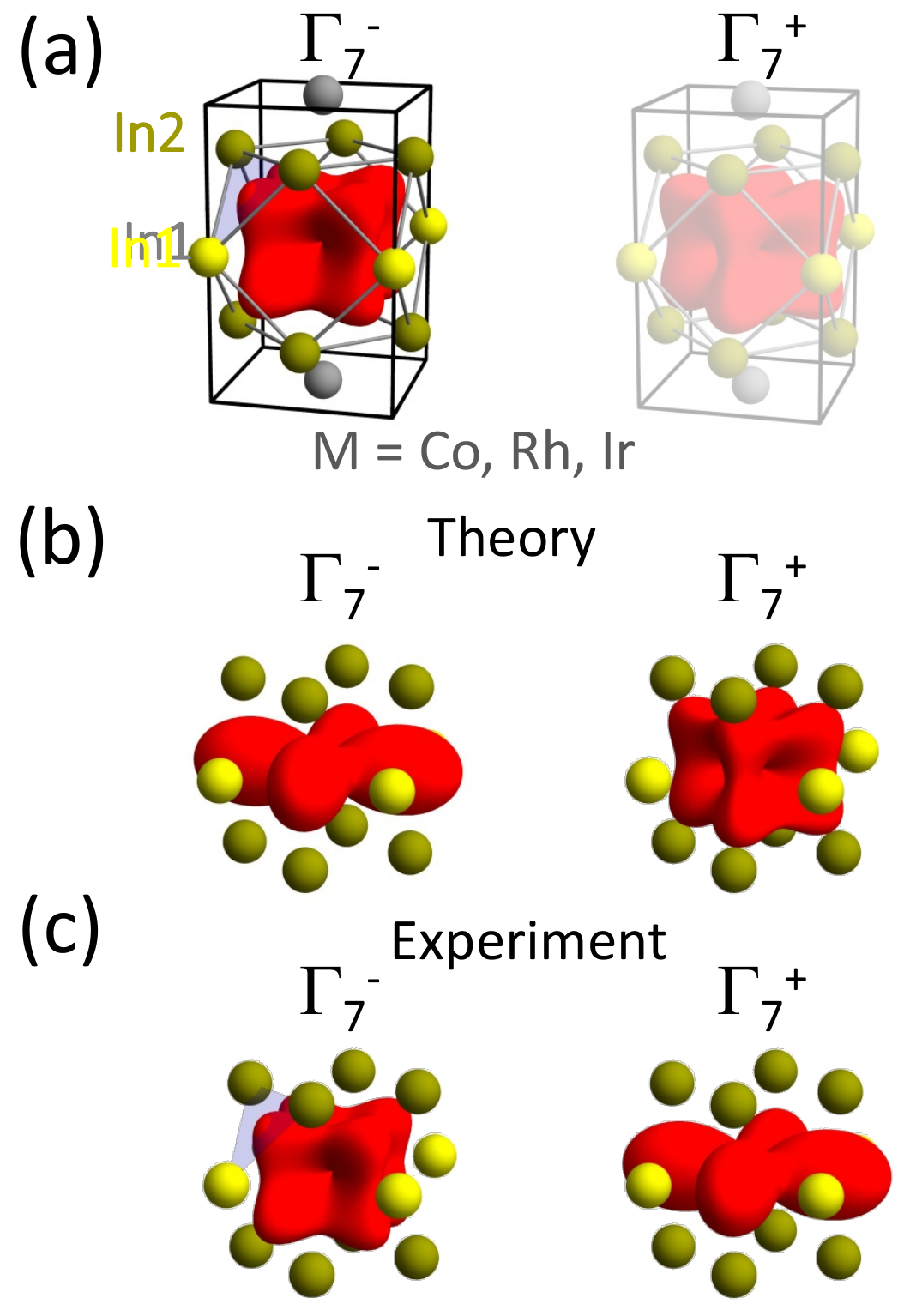}
    \caption{(color online) 
		a) Cartoon of unit cell of Ce$M$In$_5$, with orbital shape of Ce in CeCoIn$_5$ as taken from XAS for the possibility of a $\Gamma_7^-$ and $\Gamma_7^+$ Ce ground state orbital. The light blue triangle emphasizes the In2-In1-In2 triangle. 
		b) Weiss field hybridization function for Ce$M$In$_5$ from DFT+DMFT calculations decomposed into crystal-field components of the Ce ion (red orbitals) and the out-of-plane In2 (dark yellow dots) and in-plane In1 (yellow dots) environment, adapted from Ref.\,\onlinecite{Haule2010}. 
		c) Crystal-field components of the Ce ions and environment of In ions as obtained from the present NIXS experiment.
		}
    \label{fig1}
\end{figure}

A light-polarization analysis of soft X-ray absorption spectroscopy\,(XAS) spectra shows that the crystal-field wavefunction of the ground state correlates with the ground-state properties in the temperature\,-\,transition metal (substitution) phase diagram of CeCoIn$_5$\,-\,CeRhIn$_5$\,-\,CeRh$_{1-\delta}$Ir$_{\delta}$In$_5$\,-\,CeIrIn$_5$; orbitals more compressed in the tetragonal $ab$-plane favor an antiferromagnetic ground state as for CeRhIn$_5$ and the Rh rich compounds with $\delta$\,$\leq$\,0.2. The compounds with more elongated orbitals along the $c$ axis, however, have superconducting ground states (CeCoIn$_5$, CeIrIn$_5$ and also the Ir rich compounds with $\delta$\,$\geq$\,0.7).\cite{Pagliuso_2002a,Willers2015}  The obvious conclusion is that the more pronounced extension of the ground state orbitals in the direction of quantization (crystallographic $c$ direction) promotes stronger hybridization in the $z$ direction and hence superconductivity. This is supported by combined local density approximation plus dynamical mean field theory (LDA+DMFT) calculations by Shim \textsl{et al}.\,\cite{Shim2007} that find for CeIrIn$_5$ the strongest hybridization with the out-of-plane In(2) ions (see unit cell in Fig.\,\ref{fig1}\,(a)).  It was also shown that the suppression of superconductivity in CeCo(In$_{1-y}$Sn$_y$)$_5$ by about 3\% of Sn is due to a homogeneous increase of hybridization in the tetragonal $ab$ plane since the Sn ions go preferably to the In(1) sites.\cite{Sakai2015} Accordingly, we found that here the hybridization with In(1) ions plays a decisive role; the 4$f$ ground state orbital extends increasingly in the plane as the Sn content is increased.\,\cite{Chen2018}

Hence, the ground state wavefunction is a very sensitive probe for quantifying hybridization.  Haule \textsl{et al.} obtained a 4$f$ Weiss field hybridization function for Ce$M$In$_5$ based on realistic lattice parameters using density functional theory plus dynamical mean field (DFT+DMFT) calculations which they have decomposed into crystal-field components (see Fig.\,\ref{fig1}\,(b)).\cite{Haule2010} Here our goal is to verify that the crystal-field components that were extracted in these calculations are in agreement with reality. 

The tetragonal point symmetry of Ce in Ce$M$In$_5$ splits the Ce Hund's rule ground state into three Kramers doublets, two $\Gamma_7$ doublets $\Gamma_7^{+/-}$\,=\,$\left|\alpha\right||\pm5/2\rangle$\,$+/-$\,$\sqrt{1-\alpha^2}|\mp3/2 \rangle$ and $\Gamma_7^{-/+}$\,=\,$\sqrt{1-\alpha^2}|\pm5/2\rangle$\,$-/+$\,$\left|\alpha\right||\mp3/2\rangle$, and one $\Gamma_6$\,=\,$|\mp 1/2 \rangle$. We write $+/-$ or $-/+$ because the sign has not yet been determined, and this is the scope of the present manuscript.  $\Gamma_6$ as a pure $J_z$ state has full rotational symmetry around the quantization axis $c$ but the mixed states have lobes with fourfold rotational symmetry. The magnitude of $\alpha$ describes the shape and aspect ratio of the $\Gamma_7^{+/-}$ orbitals whereas the sign in the wavefunction determines how the orbitals are oriented within the unit cell; with the lobes along [100] ($\Gamma_7^+$: $x^2\,-\,y^2$) or with the lobes along [110] ($\Gamma_7^-$: $xy$). 

The crystal-field potential of the Ce$M$In$_5$ has been determined with inelastic neutron scattering (INS)\,\cite{Christianson2002,Christianson2004} and the ground state wavefunctions were studied in greater detail with linear polarized soft XAS.\,\cite{WillersPRB81,Willers2015,Chen2018}  Hence, the crystal-field energy splittings, the sequence of states ($\Gamma_7^{+/-}$, $\Gamma_7^{-/+}$, $\Gamma_6$) and also the magnitude of the $\alpha^2$-values are known (0.13, 0.38, 0.25 for Co, Rh and Ir). Only the sign of the wavefunction remains unknown because it cannot be determined with any of these dipole-selection-rule based spectroscopies. We, therefore, set up an experiment to determine the sign of the ground-state wavefunction in the Ce$M$In$_5$ compounds in order to find out which one of the two scenarios in Fig.\,\ref{fig1}\,(a) applies.  

\section{Method}
We performed a core level non-resonant inelastic x-ray scattering (NIXS) experiment at the Ce $N_{4,5}$-edges (4$d$\,$\rightarrow$\,4$f$). It has been shown previously that this method is able to detect anisotropies with higher than twofold rotational symmetry.\,\cite{Gordon2009,Willers2012,Rueff2015,Sundermann2017,Sundermann2018,Sundermann2018a} In the following, we briefly recap the principles of NIXS, a photon-in photon-out technique with hard x-rays (E$_\text{in}$\,$\approx$\,10\,keV). Because of the high incident energies, NIXS is bulk sensitive and allows one to reach large momentum transfers  $\left|\vec{q}\right|$ of the order of 10\,\AA$^{-1}$ when measuring in back scattering geometry. At such large momentum transfers, the transition operator in the scattering function S($\vec{q}$,$\omega$) can no longer be truncated after the dipole term. As a result, higher order scattering terms contribute to the scattering intensity.\,\cite{Haverkort2007,Gordon2008,Gordon2009,Bradley2010,Bradley2011,Caciuffo2010,Willers2012}  For a Ce $4d$\,$\rightarrow$\,$4f$ transition at about 10\,\AA$^{-1}$, octupole (rank $k$\,=3) and triacontadipole ($k$\,=5) terms dominate the scattering intensity whereas the dipole part ($k$\,=\,1) is less prominent. Accordingly, the directional dependence of the scattering function in a single crystal experiments follows multipole selection rules, in analogy to the dipole selection rules in linearly polarized XAS. Thus single crystal NIXS yields information not only about the orbital occupation but also the sign of the wavefunction that distinguishes the $xy$ and $x^2-y^2$ orientations of a $\Gamma_7$ when comparing two directions within the {xy}-plane; here [100] and [110].  

\begin{figure*}[t]
    \includegraphics[width=1.98\columnwidth]{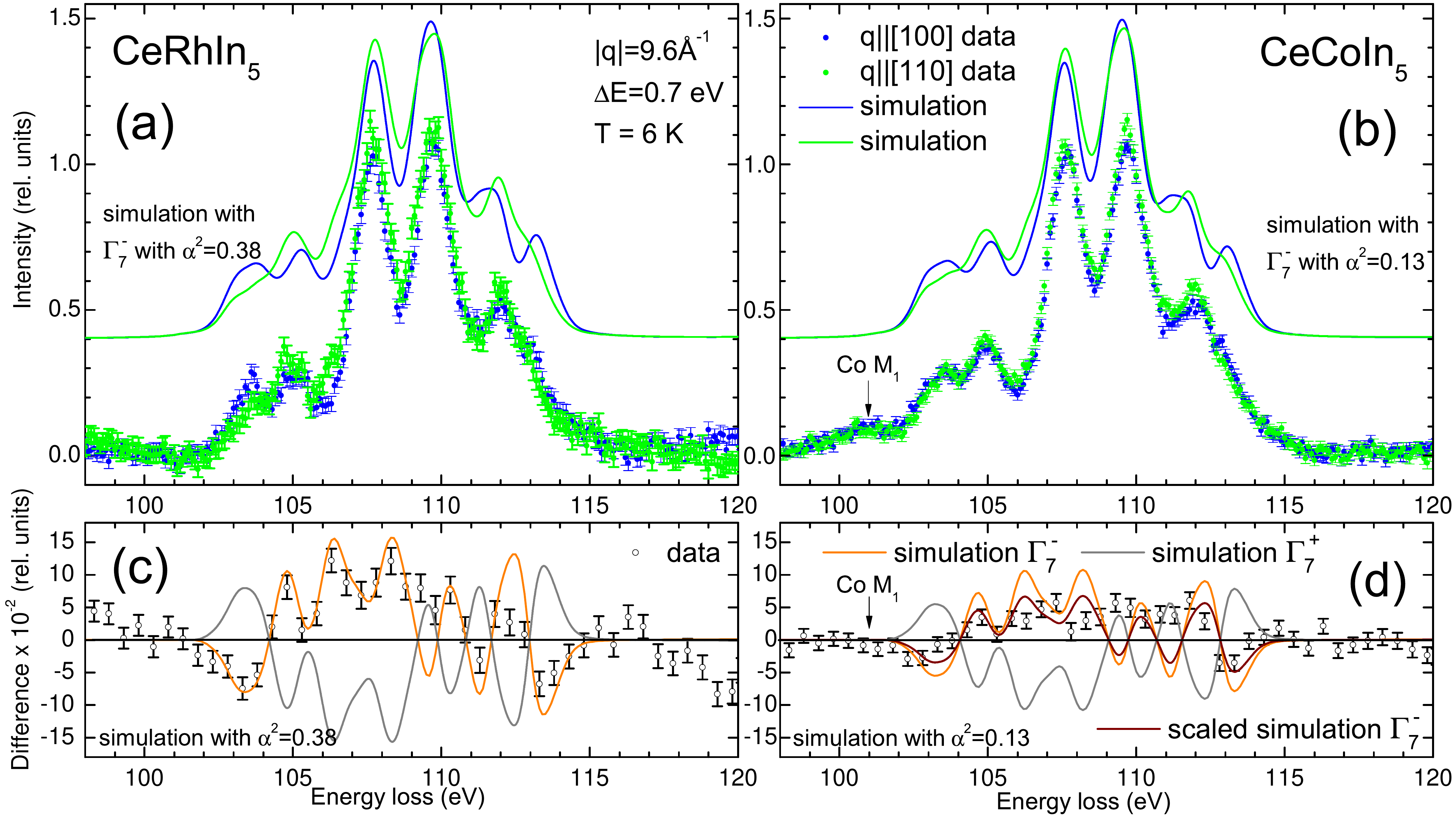}
    \caption{(color online) Non resonant inelastic x-ray scattering (NIXS) data (dots) of CeRhIn5 (a) and CeCoIn5 (b) at the Ce $N_{4,5}$-edges for the two crystallographic directions $\vec{q}$\,$\|$\,[100] (blue) and $\vec{q}$\,$\|$\,[110] (green) at $T$\,=\,6\,K, plus simulations (lines) for the respective $\Gamma_7^-$ ground states ($xy$ orientation) (see text). The bottom panels (c) and (d) show the difference spectra I$_{\vec{q}\|[110]}$\,-\,I$_{\vec{q}\|[100]}$ (circles) and simulated dichroism for the respective $\Gamma_7^-$ (orange lines) and $\Gamma_7^+$ (gray lines) crystal-field ground states, and for a mixed ground state called \textsl{scaled }$\Gamma_7^-$ (dark red lines), see text.}
    \label{fig2}
\end{figure*}

\section{Experiment}
CeCoIn$_5$ and CeRhIn$_5$ single crystals were gown using the standard In-flux technique.  CeCoIn$_5$ crystals are plate-like with the [001] direction perpendicular to the plate, whereas CeRhIn$_5$ crystals are more three-dimensional. A very detailed structural investigation on the 115 compounds shows that more than 98\,\% of the crystal volumes form in the HoCoGa$_5$ structure.\cite{Wirth2014} All samples were aligned by Laue diffraction before the experiment. For each compound two samples were cut, one with a (100) and a second one with a (110) surface so that specular geometry could be realized in the experiment.

The experiments were performed at the Max-Planck NIXS end station P01 at PETRA III/DESY in Hamburg, Germany. P01 has a vertical scattering geometry and the incident energy was selected with a Si(311) double monochromator and twelve Si(660) 1\,m radius spherically bent crystal analyzers were arranged in 3\,x\,4  array as shown in Fig.\,2 of Ref.\,\onlinecite{Sundermann2017} so that the fixed final energy was E$_\text{final}$\,=\,9690\,eV. The analyzers were positioned at scattering angles of 2\,$\theta$\,$\approx$\,150$^\circ$, 155$^\circ$, and 160$^\circ$ which provide an averaged momentum transfer of $|\vec{q}|$\,=\,9.6\,$\pm$\,0.1\,\AA$^{-1}$. The scattered beam was detected by a position sensitive custom-made detector (LAMBDA), based on a Medipix3 chip detector. The elastic line was consistently measured and a pixel-wise calibration yields instrumental energy resolutions of $\approx$\,0.7\,eV full width at half maximum (FWHM). For both samples the $N_{4,5}$ edges were measured with the momentum transfer $\vec{q}$ parallel to [001] and parallel to [110] ($\vec{q}$\,$\|$\,[001] and $\vec{q}$\,$\|$\,[110]).

We used the full multiplet code $Quanty$\,\cite{Haverkort2016} for simulating the NIXS data. A Gaussian broadening of 0.7\,eV accounts for the instrumental resolution and an additional  Lorentzian broadening of 0.4\,eV FWHM accounts for life-time effects. The atomic parameters were taken from the Cowan code,\cite{Cowan1981} whereby the Hartree-Fock values of the Slater integrals were reduced to about 60\,\% for the 4$f$-4$f$ and to about 80\,\% for the 4$d$-4$f$ Coulomb interactions to reproduce the energy distribution of the multiplet excitations of the Ce $N_{4,5}$-edges. This reduction accounts for configuration interaction processes not included in the Hartree-Fock scheme.\,\cite{Tanaka1994}

\begin{figure}[b]
    \includegraphics[width=0.99\columnwidth]{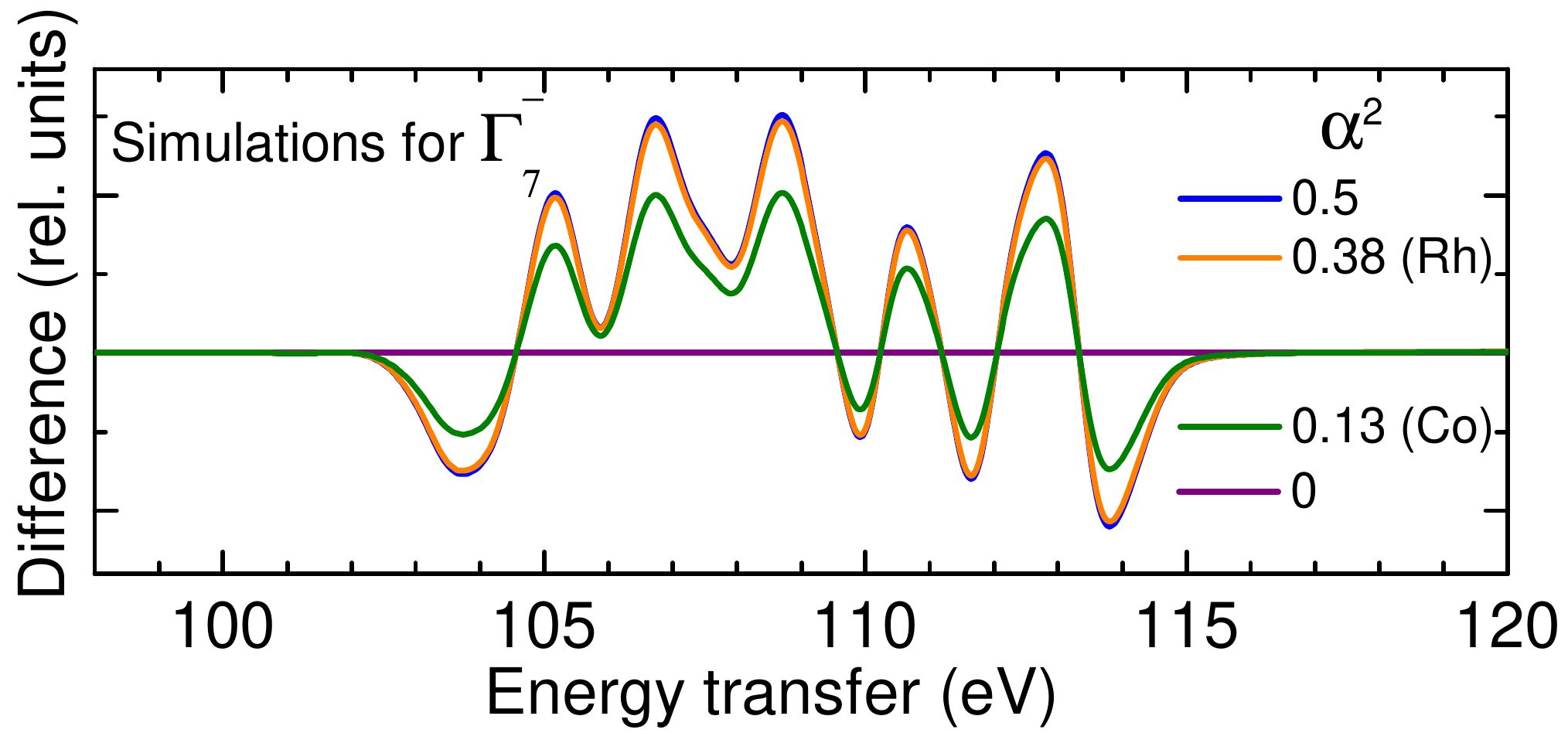}
    \caption{(color online) Simulated difference spectra I$_{\vec{q}\|[110]}$\,-\,I$_{\vec{q}\|[100]}$ for $\alpha^2$ values of 0 and 0.5, and of 0.38 for CeRhIn$_5$ and 0.13 for CeCoIn$_5$.}
    \label{fig3}
\end{figure}

\section{Results}
Figure\,\ref{fig2} shows NIXS data (circles) at the Ce $N_{4,5}$-edges of CeRhIn$_5$ (a) and CeCoIn$_5$ (b) plus simulations (lines) for two scattering directions, $\vec{q}$\,$\|$\,[100] (blue) and $\vec{q}$\,$\|$\,[110] (green). The overall shape of the spectra looks very similar and represents the multipole scattering expected for the Ce $N_{4,5}$-edges.\,\cite{Gordon2008,Gordon2009,Willers2012,Rueff2015,Sundermann2017} Figure\,\ref{fig2}\,(c) and (d) show the directional dependencies I$_{\vec{q}\|[110]}$\,-\,I$_{\vec{q}\|[100]}$ (dichroism), the experimental data as circles and simulations for the $\Gamma_7^-$ and $\Gamma_7^+$ as orange and gray lines, respectively. The expected dichroisms for a $\Gamma_7^-$ and $\Gamma_7^+$ ground state are opposite in sign so that the present experiment provides an either-or result which makes the interpretation of the data straight forward.

For CeRhIn$_5$ the $N_{4,5}$ edges in Fig.\,\ref{fig2}\,(a) as well as the dichroism in Fig.\,\ref{fig2}\,(c) are fairly well reproduced by the simulation with a $\Gamma_7^-$ ground state (orange line). Here we used the $\alpha^2$ value of 0.38 as determined with XAS.\cite{WillersPRB81} The same simulation with a $\Gamma_7^+$ ground state is clearly in contradiction to the observation (gray line). For CeCoIn$_5$ the agreement between simulated and experimental directional dependence in Fig.\,\ref{fig2}\,(d) is not as good, and we will discuss below the possible reasons for this. Also here the simulation was performed with the corresponding $\alpha^2$ value from XAS, $\alpha^2$\,=\,0.13.\cite{WillersPRB81} Most importantly, however, we conclude that the ground state of CeCoIn$_5$ must be also predominantly of $\Gamma_7^-$ character because the size of the scaling between experiment and calculation is clearly positive (0.63\,$\pm$\,0.19)\,\cite{comment} and because the $\Gamma_7^+$ is, as for CeRhIn$_5$, in clear contradiction to the observation.

\begin{figure}[]
    \includegraphics[width=0.99\columnwidth]{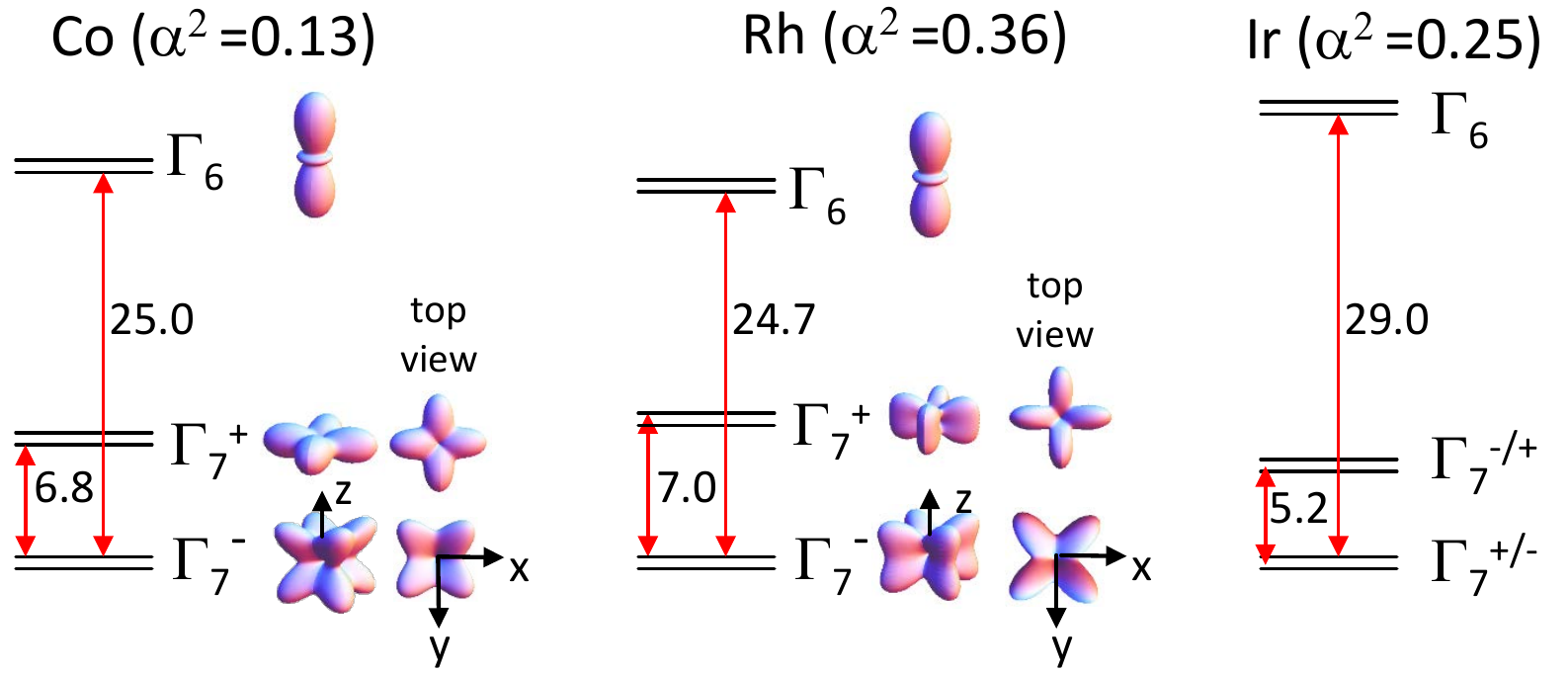}
    \caption{(color online) Crystal-field splitting of $J$\,=\,5/2 multiplet of CeCoIn$_5$, CeRhIn$_5$ and for completeness of CeIrIn$_5$ as adapted from Ref.s\,\onlinecite{Christianson2002,Christianson2004,WillersPRB81}. For $M$\,=\,Co and Rh the sign of the wavefunction is now determined which has been taken into account when drawing the $f^1$ charge densities of the respective states.}
    \label{fig4}
\end{figure}

\section{Discussion}
In Figure\,\ref{fig3}, we compare the directional dependence I$_{\vec{q}\|[110]}$\,-\,I$_{\vec{q}\|[100]}$ for several $\alpha^2$ values. For $\alpha^2$\,=\,0 (or 1) the dichroism is zero because in this case the $\Gamma_7$ state is a pure $J_z$ state and rotational invariant; for $\alpha^2$\,=\,0.5 the dichroism is largest. For CeRhIn$_5$ ($\alpha^2$\,=\,0.38) the mixing factor $\alpha^2$ is closer to 0.5 than for CeCoIn$_5$ ($\alpha^2$\,=\,0.13) so that the expected dichroism for CeCoIn$_5$ is smaller than for CeRhIn$_5$. But, the expected reduction due to the different $\alpha^2$ values still does not account for the strongly reduced directional effect in CeCoIn$_5$. A natural explanation could be the stronger $cf$-hybridization in CeCoIn$_5$ with respect to CeRhIn$_5$: in CeCoIn$_5$ the coherence temperature $T^*$ is of the order of 50\,K\cite{Petrovic2001,WillersPRB81} which is comparable to the energy splitting of the two lowest crystal-field states (6.8\,meV or $\approx$75\,K) (see Fig.\,\ref{fig4}) so that the first excited crystal-field state will contribute to the ground state via hybridization. The first excited crystal-field state is the $\Gamma_7^+$ which has the opposite dichroism of the crystal-field ground state $\Gamma_7^-$ so that the net dichroism of the hybridized ground state will be reduced. In short, the strongly reduced directional effect in CeCoIn$_5$ is due to the presence of strong hybridization. Assuming the first excited crystal-field state $\Gamma_7^+$ contributes 19\,\% to the ground state of CeCoIn$_5$ yields a very good agreement of measured dichrosim and simulation (see dark red line in Fig.\,\ref{fig2}\,(d)).\cite{comment} However, 19\,\% of $\Gamma_7^+$ mixed into the ground state by hybridization must be an overestimation because the $\Gamma_7^+$ admixture was not accounted for when describing the linear dichroism in XAS.\cite{WillersPRB81} It turns out that both data sets, the directional dependence in NIXS and the linear dichroism in XAS, can be analyzed consistently and are well described with $\alpha^2$\,=\,0.10 and 13\,\% of $\Gamma_7^+$.

Figure\,\ref{fig4} summarizes what we know now about the crystal-field splittings of the $J$\,=\,5/2 multiplet of Ce$M$In$_5$. The splittings and $\alpha^2$ values are taken from inelastic neutron scattering and XAS as published in Ref.\,\onlinecite{Christianson2002,Christianson2004,WillersPRB81}. The present NIXS experiments on CeCoIn$_5$ and CeRhIn$_5$ add the missing information that the $\Gamma_7^-$ dominates the ground state for both compounds, i.e.\,the lobes of the crystal-field ground state are along the crystallographic (110) direction.  These $\Gamma_7^-$ ground state orbitals extend more in the $z$-direction than the $\Gamma_7^+$ at about 6-7\,meV so that the scenario as shown in Fig.\ref{fig1}\,(c) applies to CeCoIn$_5$ and CeRhIn$_5$, whereas the wavefunctions projected out by DFT+DMFT calculations\cite{Haule2010} have opposite signs (see Fig.\ref{fig1}\,(b)). 

The tips of the lobes of the $\Gamma_7^-$ ground state wavefunctions of CeCoIn$_5$ and CeRhIn$_5$ are pointing towards the triangle In2-In1-In2 (see Fig.\,\ref{fig1} or Fig.\ref{fig1}\,(c)). It is therefore reasonable to conclude that the impact of the hybridization with the out-of-plane In2 atoms is more important than the hybridization with the in-plane In1 atoms. This is in agreement with the results of the CeRh$_{1-\delta}$Ir$_{\delta}$In$_5$ substitution series\,\cite{Willers2015}  where the orbitals that are more extended along the $c$-axis tend to hybridize more strongly. Nevertheless, the present results also show that hybridization with the In1 atoms is important and this supports the results of the CeCo(In$_{1-y}$Sn$_y$)$_5$ substitution series;\cite{Chen2018} the Sn atoms go preferably to the In1 sites leading to a stronger hybridization in the plane.\cite{Sakai2015} \red{We would like to note that the revised value of $\alpha^2$ for CeCoIn$_5$ that is obtained when taking into account the first excited crystal-field state leads to a crystal-field ground state orbital that is even more extended in the $z$-direction than for the originally anticipated value. The same should apply to CeIrIn$_5$ when allowing a hybridization induced contribution of the first excited crystal-field state.  Hence, the correlation of stronger hybridization with the In2 atoms due to ground state orbitals that are more extended in $z$-direction and superconductivity still holds.\cite{Willers2015} }

A modest increase in the contribution of $J_z$\,=\,$|\pm5/2\rangle$ to the $\Gamma_7^-$-state of CeRhIn$_5$ will promote overlap of $f$-orbitals with $p$-states of In(1) at the expense $f$-In(2) hybridization. Mixing of Zeeman-split ground and $\Gamma_7^+$ first excited crystal-field levels, in principle, could produce such a modest increase in the $J_z$\,=\,$|\pm5/2\rangle$ contribution.  Indeed, recent high-field magnetostriction \cite{Rosa2019} and nuclear magnetic resonance \cite{Lesseux2018} measurements on CeRhIn$_5$ are consistent with this possibility that appears to be a significant contributing factor to field-induced Fermi-surface reconstruction in CeRhIn$_5$ subject to a magnetic field near 30\,T. 

In the limit of strong intra-atomic Coulomb interactions, which is typical of strongly correlated metals like CeCoIn$_5$ and CeRhIn$_5$, the magnetic exchange $\cal{J}$ is proportional to the square of the matrix element $\left\langle V_{kf}\right\rangle$ that mixes conduction and $f$-wavefunctions.\,\cite{Schrieffer1966} Both Kondo and long-range Ruderman-Kittel-Kasuya-Yosida (RKKY) interactions depend on the magnitude of $\cal{J}$ that is set by $\left\langle V_{kf}\right\rangle$$^2$ and, consequently, by the $f$-orbital configuration.  These interactions are fundamental for a description of Kondo-lattice systems and their relative balance can be tuned by non-thermal control parameters, such as magnetic field and pressure. Modest pressure applied to CeRhIn$_5$ tunes its antiferromagnetic transition temperature toward zero temperature where a dome of unconventional superconductivity emerges with a maximum transition temperature very close to that of CeCoIn$_5$\,\cite{Thompson2012} and also changes the Fermi surface from small to large as in CeCoIn$_5$.\cite{Shishido2005} We do not know if the $f$-orbital configuration of CeRhIn$_5$ at these pressures is the same as that of CeCoIn$_5$, but this is an interesting possibility that merits study.

\section{Summary}
In $f$-based materials, the shape of the crystal-field wavefunctions ultimately determines the origin of anisotropic hybridization in these materials and their ground state. Here, we show that the ground state of Ce$M$In$_5$ ($M$\,=\,Co,Rh) is a $\Gamma_7^-$\,=\,$\left|\alpha\right||\pm5/2\rangle$\,$-$\,$\sqrt{1-\alpha^2}|\mp3/2 \rangle$ doublet with lobes $\Gamma_7^-$ pointing toward the 110 direction, i.e., the lobes have $xy$  character. Though careful DFT+DMFT calculations shed light on these materials, the crystal-field scheme obtained is different from our experimental one. Our work settles the question on the orientation of $f$-orbitals in the ground state of Ce$M$In$_5$ and will stimulate theoretical developments that take into account the actual wavefunctions. 

\section{Acknowledgment} We thank Peter Thalmeier for fruitful discussions. Parts of this research were carried out at PETRA\,III/DESY, a member of the Helmholtz Association HGF, and we would like to thank Christoph Becker, Manuel Harder and Frank-Uwe Dill for skillful technical assistance. Work at Los Alamos was performed under the auspices of the U.S. Department of Energy, Office of Basic Energy Sciences, Division of Materials Science and Engineering.  A.A., A.S., and M.S. gratefully acknowledge the financial support of the Deutsche Forschungsgemeinschaft under projects SE\,1441-4-1.


\begin{thebibliography}{42}
\bibitem [{com()}]{comment}%
  \BibitemOpen
  \href@noop {} {\ }\bibinfo {note} {{The values and standard errors of the
  mean of the experimental dichroism weighted by and normalized to the one of
  the $\Gamma_7^{-}$ simulation amount to ($0.60 \pm 0.07$) and ($0.18 \pm
  0.05$) for CeRhIn$_5$ and CeCoIn$_5$, respectively. We can estimate the
  scaling between experiment and theory from these values by dividing by the
  Pearson correlation coefficients $c$ = 0.72 for CeRhIn$_5$ and 0.28 for
  CeCoIn$_5$ and obtain ($0.83 \pm 0.10$) for CeRhIn$_5$ and ($0.63 \pm 0.19$)
  for CeCoIn$_5$.}}\BibitemShut {Stop}%
\end{thebibliography}
\end{document}